\documentclass[
    ,final            
  ]
  {aipproc}

\layoutstyle{8x11double}
\def\lsim{\raise0.3ex\hbox{$<$\kern-0.75em\raise-1.1ex\hbox{$\sim$}}}

\begin{document}

\title{Universal critical behavior and the transition temperature 
  in (2+1)-flavor QCD}


\author{Christian Schmidt $\;${\it for the HotQCD Collaboration}
  \footnote{HotQCD Collaboration members are: A.~Bazavov,
    T.~Bhattacharya, M.~Cheng, N.H.~Christ, C.~DeTar, S.~Gottlieb,
    R.~Gupta, P.~Hegde, U.M.~Heller, C.~Jung, F.~Karsch, E.~Laermann,
    L.~Levkova, C.~Miao, R.D.~Mawhinney, S.~Mukherjee, P.~Petreczky,
    D.~Renfrew, C.~Schmidt, R.A.~Soltz, W.~Soeldner, R.~Sugar,
    D.~Toussaint, W.~Unger, P.~Vranas}}{
  address={Frankfurt Institute for Advanced Studies,  
    J.W.Goethe Universit\"at Frankfurt, D-60438 Frankfurt, Germany}
  ,altaddress={GSI Helmholtzzentrum f\"ur Schwerionenforschung, 
    Planckstr.~1, D-64291 Darmstadt, Germany}
}

\begin{abstract}
  We discuss the universal critical behavior in (2+1)-flavor QCD by
  analyzing lattice data from improved staggered fermions generated by
  the HotQCD Collaboration. We present recent results from two
  different lattice discretizations and various lattice spacings
  ($N_\tau=6,8,12$) at fixed physical strange quark mass ($m_s$) but
  varying light quark mass ($m_l$). We find that the chiral
  order-parameter, i.e. the chiral condensate, shows the expected
  universal scaling that is associated with the critical point in the
  chiral limit already for light quark masses $m_l/m_s \lsim 0.05$.
  From an analysis of the disconnected chiral susceptibility we
  estimate a preliminary value of the QCD transition temperature.
\end{abstract}

\maketitle


\section{Introduction}
For almost 3 decades, lattice QCD is providing us with valuable
information on QCD thermodynamics. Only recently, however, the
numerical calculations arrive at a precision level, where controlled
continuum extrapolations become possible. Moreover, our understanding
of lattice cut-off effects resulting from the discretization of the QCD
Lagrangian developed over the years. 

We will present here results from two improved staggered actions:
asqtad, which has been widely used for finite temperature calculations
in the past \cite{DeTar} and the recently introduced highly improved
staggered quark (HISQ) action \cite{HISQ}.  Apart from different
treatments in the gauge part, these actions mainly differ in their
strategies for the reduction of the flavor symmetry breaking, which is
inherent to the staggered fermion formulation.  The HISQ action has
been shown to offer the most superior degree of improvement in the
flavor symmetry sector \cite{HISQ,MILC_HISQ}.  In the following we will
refer to the action that incorporates the HISQ improvement program in
the fermion sector together with a tree-level Symanzik improvement in
the gauge part as HISQ/tree.

At sufficiently high temperature, QCD undergoes a transition form
hadronic to quark degrees of freedom, which is a true phase transition
only in the limit of massless quarks. This transition, driven by the
restauration of the chiral symmetry, is of second order in the chiral
limit. All thermodynamic quantities that can be derived as derivatives
from the QCD partition function are thus subject to a universal
critical behavior.  In the following we will analyze the universal
critical behavior of the chiral condensate, which is a true
order-parameter in the chiral limit. Furthermore we estimate the
transition temperature from the disconnected part of the chiral
susceptibility.

The lattice data has been generated by the HotQCD collaboration
\cite{HotQCD} with the above mentioned actions on lattices with
temporal extent of $N_\tau=6,8$ and for the asqtad action with 
$N_\tau=12$. This translates into
lattice spacings $a$ of roughly $a=0.16,0.12$ and $0.08$~fm (at
$T=200$ MeV). Throughout all calculations the strange quark mass $m_s$
was fixed to its physical value, whereas we have varied the light
quark mass $m_l$ in the range $m_l/m_s=0.05-0.2$. The physical value
of light to strange quark mass is approximately given by
$m_l/m_s=0.037$.

\begin{figure}
  \includegraphics[height=.35\textwidth]{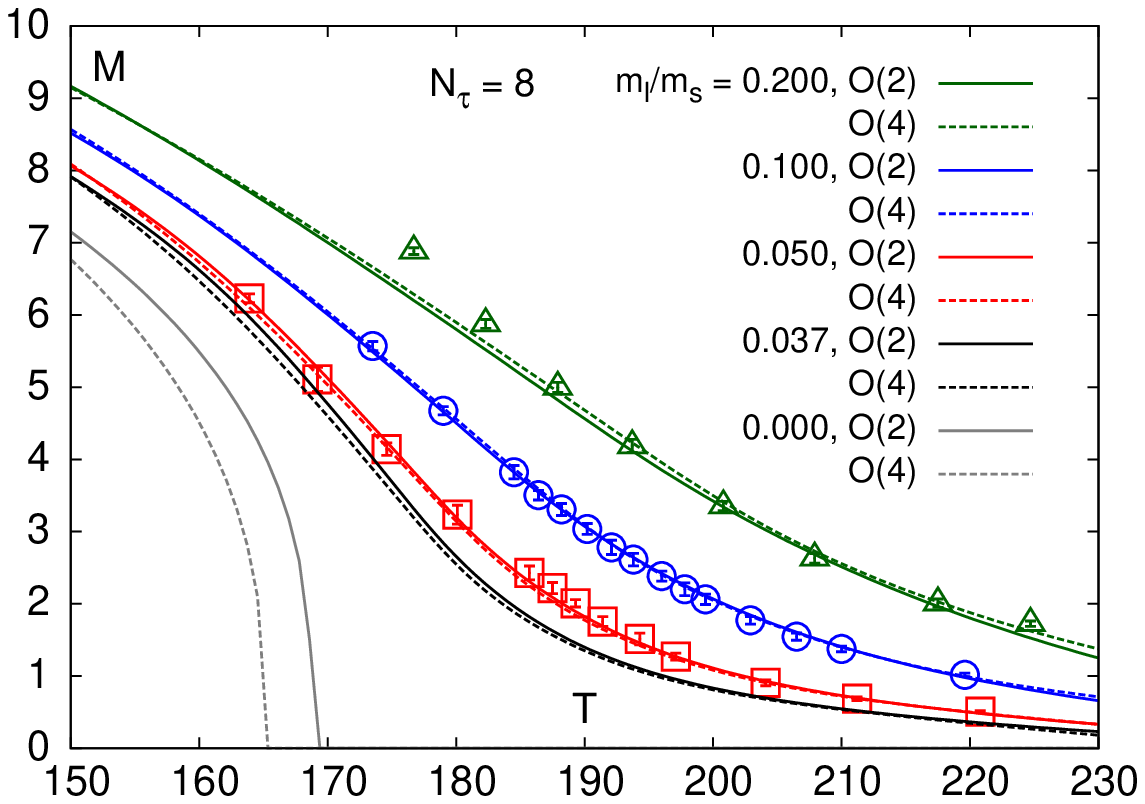}
  \includegraphics[height=.35\textwidth]{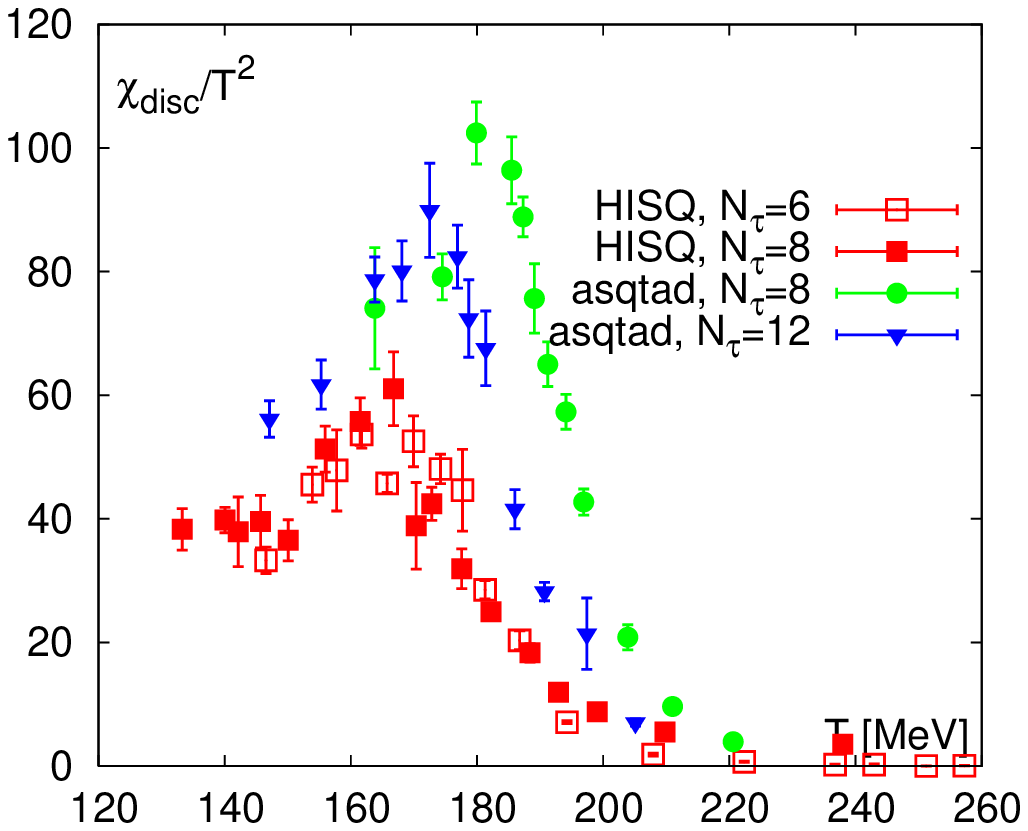}
  \caption{Left: Chiral order-parameter as defined in Eq.~(\ref{eq:M})
    for different quark masses, plotted as a function of
    temperature. Simulations have been performed with the asqtad
    action obtained on $N_\tau=8$ lattices. Right: Disconnected part of
    the chiral susceptibility for $m_l/m_s=0.05$. Compared are results
    from two different lattice actions and different lattice
    spacings.\label{fig:chi}}
\end{figure}

\section{Universal scaling}
The QCD phase transition in the chiral limit of 2-flavor QCD is
expected to be in the $O(4)$ universality class \cite{Uclass}. For a
sufficiently large strange quark mass this is expected to be the same
in (2+1)-flavor QCD. The staggered fermion formulation on the lattice brakes
chiral symmetry. For each fixed lattice spacing one is only 
left with a $O(2)$ symmetry. The $O(4)$ symmetry will, of course, be 
restored in the continuum limit. 

In order to analyze the universal critical behavior that arises
in the chiral limit we have to arrange the free energy (and its
derivatives) in terms of the relevant scaling variables.
In general one separates two kinds of contributions to the free energy
density, a part ($f_s$) that will generate singularities in higher
order derivatives of the partition function and a regular part
($f_r$), we define
\begin{equation}
f(T,m_l,m_s)=f_s(T,m_l,m_s)+f_r(T,m_l,m_s)\;.
\end{equation}
Although $f_s$ depends on many parameters of the QCD Lagrangian, the
universal behavior can be expressed in only two relevant scaling
variables t and h, that control deviations from criticality at $(t, h)
= (0, 0)$. The scaling variables are up to some normalization factors
$t_0$, $h_0$ given by the reduced temperature $t=(T-T_c)/(t_0 T_c)$
and the symmetry braking parameter, which we will define here as
$h=m_l/(h_0 m_s)$.  The universal critical behavior of the order
parameter, $M \sim \partial f / \partial m_l$ , is controlled by a
scaling function $f_G$ that arises from the singular part of the free
energy density after taking a derivative with respect to the light
quark mass. We make use of the fact that $f_s$ is a homogeneous function
of its arguments. $f_G$ (and $f_s$) can thus be written as a function of 
the single scaling variable $z = t /h^{1/\delta \betaδ}$, where $\delta$ 
and $\beta$ are
critical exponents of the three dimensional $O(N)$ universality
class. In addition we also consider the leading oder contribution
coming from the regular term of the free energy. We finally arrive at
\begin{equation}
M=h^{1/\delta}f_G(z)+a_tth+b_1h
\label{eq:scaling}
\end{equation}
The scaling function $f_G$ is well known from spin-model simulations
and easily accessible via, e.g., the implicit parameterization given in 
\cite{fG}. On the right hand site of Eq.~(\ref{eq:scaling}) we are
left with 5 unknown parameters, $t_0,h_0,T_c,a_t,b_1$, which have to 
be determined through a fit to the lattice data. Out of these,
$T_c$ is the only one that does not depend on the particular definition
of the order-parameter $M$. All of them are, however, cut-off dependent
and needed to be extrapolated to the continuum limit.

On the lattice the chiral order parameter (chiral condensate) is always 
finite but contains contributions that diverge in the continuum limit, 
{\it i.e.} it requires renormalization and in particular an additive
and multiplicative renormalization. In order to remove at least the
multiplicative renomalization factor already on the lattice we consider
here the following definition of the order-parameter
\begin{equation}
M=m_s\left<\bar\psi\psi\right>_l/T^4\;.
\label{eq:M}
\end{equation}
In Fig.~\ref{fig:chi} (left) we show the chiral oder-parameter as
defined in Eq.~(\ref{eq:M}) from simulations with the asqtad action
obtained on lattices with temporal extend $N_\tau=8$ and light quark
masses ranging from $m_l/m_s=0.05-0.2$. Also shown are fits to the
$O(2)$ (solid curves) and $O(4)$ (dashed curves) scaling functions. 
In the fits we have omitted the data set with the largest quark mass 
$m_l/m_s=0.2$. 
For masses $m_l/m_s\lsim 0.1$ the fit describes
the lattice data rather good in the entire temperature range we have
considered.  This coincides with finding obtained with the p4 action
on $N_\tau=4$ \cite{magnetic} and $N_\tau=8$
\cite{curvature}. Using the fits, we can extrapolate the chiral
order-parameter to the physical mass ($m_l/m_s=0.037$) and the chiral
limit, as also indicated in the Fig.~\ref{fig:chi} (left). The main 
difference between the fits manifests in the chiral limit. We 
obtain values for $T_c$ranging from $T_c\approx (165-170)$~MeV (obtained 
for $N_\tau=8$ with the asqtad action).

\begin{figure}
  \includegraphics[height=.33\textwidth]{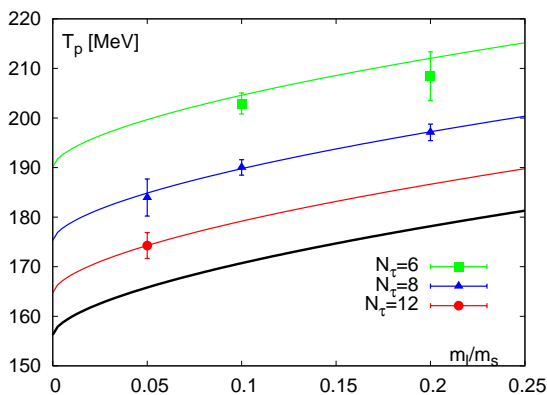}
  \caption{Transition temperature as obtained from the peak position
    of the disconnected part of the chiral susceptibility for different
    light quark masses and lattice spacings from simulations with the 
    asqtad action. Also show is a combined
    chiral and continuum fit with the Ansatz given in Eq.~\ref{eq:Tp}.
    \label{fig:Tp}}
\end{figure}

\section{The critical temperature}
It is the ultimate goal and a valuable input for many phenomenological
calculations, e.g. in heavy ion phenomenology, to obtain a continuum
extrapolated value for $T_c$. The chiral susceptibility is
traditionally used to determine the QCD transition temperature. This
quantity exhibits a peak as plotted as a function of temperature,
which position can be associated with the QCD transition.  Note that
at nonzero light quark mass the QCD transition is a smooth crossover
rather then a genuine phase transition of second Order.  In this case
the corresponding pseudo-critical temperature is not unique; it
depends on the observable used to define it.  All definitions will,
however, lead to the same value for $T_c$ in the chiral and continuum
limit.

The chiral susceptibility which has a connected and disconnected part
is defined as the mass derivative of the order-parameter 
\begin{equation}
\chi\equiv \frac{T}{V}\frac{\partial }{\partial m_l}
\left<\bar\psi\psi\right>_l\equiv \chi_{\rm dis}+\chi_{\rm con}\;.
\end{equation}
In Fig.~\ref{fig:chi} (right) we plot the disconnected part of the 
chiral susceptibility as obtianed for light quark mass $m_l/m_s=0.05$
but different lattice spacings. We compare results from two different 
lattice actions: asqtad and HISQ/tree. We find that the cut-off effects
are visible in the peak hight and peak position. They are considerably 
smaller in case of the HISQ/tree action. 

From the scaling function for the chiral condensate,
Eq.~(\ref{eq:scaling}), the scaling of the chiral susceptibility can
be easily extracted. It is governed by the function $f'_G(z)\equiv
{\rm d}f_G/{\rm d}z$. To obtain a consistent picture it is desirable
to perform a similar analysis as presented for the order-parameter in
the previous section. This should lead to the same non-universal
normalizations of the scaling fields as well as to the same value for
$T_c$ in the chiral limit. At the moment, however, we restrict ourselves 
to an analysis of the peak positions $T_p$ of the disconnected
part $\chi_{\rm dis}$. From Eq.~(\ref{eq:scaling}) we obtain the fit
Ansatz
\begin{equation}
T_p(m_l,N_\tau)=T_c+b(m_l/m_s)^d+cN_\tau^{-2}\;.
\label{eq:Tp}
\end{equation}
Here we allow for a generic cut-off dependence of the form $1/N_\tau^2$ 
in addition to the mass dependence coming from the scaling Ansatz. We thus 
have three fit parameters, $T_c$, $b$, $c$. The exponent $d$ is fixed by
universal scaling as $d=1/\beta\delta$.

In Fig.~\ref{fig:Tp} we show the peak positions $T_p$ from $\chi_{\rm dis}$ as 
obtained by simulations with the asqtad action for different quark masses
and lattice spacings. The values and their errors have been estimated
by two different asymmetric fits to the peak \cite{wolfgang}. In addition 
we show a combined chiral and continuum extrapolation based on 
Eq.~(\ref{eq:Tp}). The fit works reasonably well and we obtain 
as a preliminary result for the transition temperature at physical masses: 
\begin{equation}
T_p(0.037m_s,\infty)=164(6)\;\mbox{MeV.}
\end{equation} 
Here the error summarizes our estimate of the statistical and systematical 
error.


\begin{theacknowledgments}
This work has been supported in part by contracts DE-AC02-98CH10886
and DE-FC02-06ER-41439 with the U.S. Department of Energy and contract
0555397 with the National Science Foundation. The numerical
calculations have been performed using USQCD resources at Fermilab and
JLab, the BlueGene/L at the New York Center for Computational Sciences
(NYCCS), and the BlueGene/L at the J\"ulich Supercomputing Center. CS has 
partially been supported through the Helmholtz International Center for 
FAIR which is part of the Hessian LOEWE initiative.
\end{theacknowledgments}

\end{document}